\documentclass{ptephy}

\newcommand{\beq}{\begin{eqnarray}}
\newcommand{\eeq}{\end{eqnarray}}

\usepackage{color}

\begin{document}

\title{Phase structure of pure SU($3$) lattice gauge theory in $5$ dimensions}

\author{\name{\fname{Etsuko} \surname{Itou}}{1,\ast},
\name{\fname{Kouji} \surname{Kashiwa}}{2}, and
\name{\fname{Norihiro} \surname{Nakamoto}}{1,3}}

\address{
\affil{1}{High Energy Accelerator Research Organization (KEK), Tsukuba 305-0801, Japan}
\affil{2}{RIKEN/BNL, Brookhaven National Laboratory, Upton, New York 11973, USA}
\affil{3}{Kanazawa University, Institute for Theoretical Physics (KITP) Kakuma-machi, Kanazawa 920-1192 JAPAN}
\email{eitou@post.kek.jp}
}

\begin{abstract}%
We investigate the nonperturbative phase structure of five-dimensional SU($3$) pure Yang-Mills theory on the lattice.
We perform numerical simulations using the Wilson plaquette gauge action on an anisotropic lattice with a  four-dimensional lattice spacing ($a_4$) and with an independent value in the fifth dimension ($a_5$).
We investigate both cases of $a_4 > a_5$ and $a_4 < a_5$.
The Polyakov loops in the fourth and the fifth directions are observed, and we find that there are four possible phases for the anisotropic five-dimensional quenched QCD theory on the lattice.
We determine the critical values of the lattice bare coupling and the anisotropy parameter for each phase transition.
Furthermore, we find that the two center domains where the Polyakov loop have locally different charges of the center symmetry appears in single configuration in the specific region of the lattice parameters.
\end{abstract}

\subjectindex{B01, B43}

\maketitle

\section{Introduction}
Lattice gauge theory is one of the regularization methods for the quantum gauge theories.
Moreover, it is the only known regularization method which respects the gauge invariance.
The regulator is introduced as a lattice spacing ($a$), and the frequency modes higher than ($1/a$) are suppressed because of the ultraviolet (UV) cutoff.
In the four-dimensional non-abelian gauge theory, the lattice numerical simulation in the strong coupling limit smoothly connects to the weak coupling limit, so that we can reach the nonperturbative regime from the well-defined continuum limit ($a \rightarrow 0$) which corresponds to the lattice bare coupling goes to zero.
The numerical simulation based on the lattice QCD have reproduced the nonperturbative dynamics of the quarks and the gluons.

The five-dimensional SU($2$) gauge theory with the Wilson plaquette gauge~\cite{Creutz:1979dw} or the mixed gauge action~\cite{Kawai:1992um} have been investigated.
The difference from the four-dimensional SU($2$) gauge theory is that the numerical simulation exhibits a clear hysteresis at a finite lattice bare coupling.
The phase transition is known as the bulk first order transition coming from lattice artifacts.
In such theory, we cannot give a definition of the continuum limit in whole region of the coupling constant.
A similar hysteresis is also observed in the four-dimensional abelian gauge theory~\cite{Creutz:1979dw}.
The theory is well known to be ill-defined in the high energy region, and only the perturbative picture is allowed at the low energy scale.
Based on the analogy with the four-dimensional abelian theory, the existence of the first order bulk phase transition might suggest that the higher dimensional gauge theory would be also an ill-defined theory in the nonperturbative sense.

\begin{figure}[h]
\vspace{2cm}
\begin{center}
\includegraphics[scale=0.3]{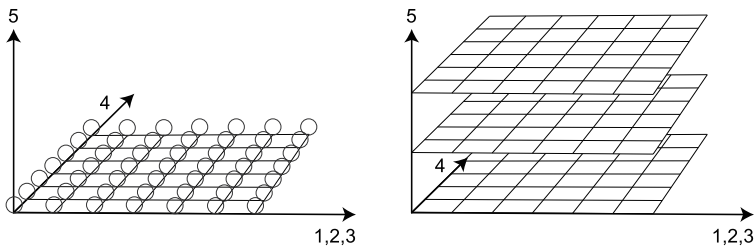}
\caption{Two possible pictures of the anisotropic five-dimensional lattice geometry. In the left panel, we assume that the lattice spacing of the fifth dimension is smaller than the one of the other directions.
On the other hand, the lattice spacing of the fifth dimension is larger than the one of the other directions in the right panel.}
\label{fig:image}
\end{center}
\end{figure}
On the other hand, even if the theory does not have a well-defined continuum limit as a whole five-dimensional theory,
it is worth to consider the effective theory which has an explicit finite UV cutoff only in the fifth dimension, and to estimate the nonperturbative contributions to the four-dimensional Yang-Mills theory in the continuum limit as an extended model.

There are two independent ideas to construct a four-dimensional effective model starting from the five-dimensional gauge theory.
The first one has the geometry whose fifth dimension is compact space as shown the left panel in Fig.~\ref{fig:image}.
We can observe only the low energy physics in four dimensions  when the inverse of the compactification radius in the extra dimension is smaller than the observation scale.
Several phenomenological models try to explain the origin of the Higgs particle \cite{Csaki:2005vy, Manton:1979kb, Fairlie:1979at, Hosotani:1983xw, Hosotani:1988bm} or the candidates of the dark matter \cite{Hosotani:2009jk} by the Kaluza-Klein (KK) modes coming from the compactification.
The other geometrical picture is naively given by the right panel in Fig.~\ref{fig:image}.
If the four-dimensional branes weakly interact with each other~\cite{Fu:1984gj}, then we can not observe a contribution from the other four-dimensional branes even if they exist in the extra dimension. 
Several phenomenological models are proposed based on Arkani-Hamed-Dimopoulos-Dvali (ADD)~\cite{ArkaniHamed:1998rs}
 or Randall-Sundrum~\cite{Randall:1999ee,Randall:1999vf} model in which the gauge and matter fields are localized on the four-dimensional brane.
However, whether such localization occurs or not is an independent problem.
It is important to find the phase where the quark currents do not extent to the fifth direction, although the four-dimensional dynamics on the brane shows the ordinary QCD.

In this work, we investigate the phase structure of the five-dimensional pure SU($3$) gauge theory with three geometrical pictures, namely the isotropic one and both anisotropic pictures shown in Fig.\ref{fig:image}.
This is the first study on the five-dimensional SU($3$) gauge theory on the lattice, although there are several works for the SU($2$) gauge theories~\cite{Creutz:1979dw, Kawai:1992um, Ejiri:2000fc}--\cite{Nicolis:2010nh}.
The SU($3$) gauge theory has a higher rank of the group than the SU($2$) gauge theory.
We expect that the SU($3$) gauge theory has a richer phase structure in comparison with the SU($2$) gauge theory.
We mainly observe the Polyakov loop on the lattice for more than $2,000$ lattice parameters, and find that there are in total four possible phases for the Wilson plaquette gauge action with an anisotropy parameter which describes the ratio of the lattice spacings between the fourth and the fifth directions.
We determine the critical values of the lattice bare gauge coupling constant and the anisotropy parameter for each phase transition.

This paper is organized as follows.
We give a definition of the model for the pure SU($3$) lattice gauge theory given by the Wilson plaquette gauge action, and show the conjectured phase diagram of the model in Sec.~\ref{sec:definition}.
We investigate the phase structure of the isotropic SU($3$) lattice gauge theory in five dimensions in Sec.~\ref{sec:iso-lattice}, and find an evidence of the hysteresis.
It indicates that there is a bulk first order phase transition which is also observed in the five-dimensional SU($2$) lattice gauge theory.
In Sec.~\ref{sec:aniso-lattice}, we introduce an anisotropy between the fourth- and the fifth-dimensional lattice spacings to realize rich geometrical pictures.
The phase transition points, where the strong hysteresis disappears, exist in the small and the large anisotropy parameter regions.
We determine the critical values of the lattice bare gauge coupling constant for the spontaneous center symmetry breaking and the anisotropy parameter where the strong hysteresis disappears for several lattice sizes.
In Sec~\ref{sec:meta-stable}, we report that the appearance of the kink configurations which has a domain-wall of the two different charges of the center symmetry~\cite{Asakawa:2012yv}.
Thus, the Polyakov loop with the two different $Z_3$ charges coexists in such configurations.

\section{The model and simulation setup}\label{sec:definition}
\subsection{The action}
The SU($N_c$) Yang-Mills gauge action in five dimensions can be given by
\beq
S=\int d^4x \int dx_5 \frac{1}{2} \mathrm{Tr} F_{MN}^2,\label{eq:cont-action}
\eeq
where $M,N=1, \cdots , 5$ and $F_{MN}= \partial_M A_N -\partial_N A_M + ig_5 [ A_M, A_N ]$.
Here $g_5$ is a five-dimensional gauge coupling constant and the mass dimension of $g_5$ is $-1/2$.
Therefore this theory is nonrenormalizable at least within the perturbative expansion.

To regulate the theory, we introduce a finite lattice spacing ($a$), which naively corresponds to a inverse of the UV cutoff.
On the lattice, one of possible actions is the Wilson plaquette gauge action,
\beq
S&=& \beta \sum_{x} \sum_{1 \le M \le N \le5} \left( 1-\frac{1}{N_c} \mathrm{Tr} U_{M N} \right),\label{eq:iso-action}
\eeq
where $\beta$ is a normalization factor, which we will determine to reproduce the continuum action (Eq.~(\ref{eq:cont-action})).
Here $U_{M N}$ and $U_M (x)$ denote the plaquette and the link variable given by
\beq
U_{M N}&=&U_{M}(x) U_{N}(x+\hat{M}a) U^\dag_M (x+\hat{N}a)U^\dag_N (x),\\
U_M (x)&=&e^{i a g_5 A_M (x+\hat{M}/2)},\label{eq:link-variable}
\eeq
respectively.
Substituting Eq.~(\ref{eq:link-variable}) to the lattice action (Eq.~(\ref{eq:iso-action})), we obtain the following action in the naive $a \rightarrow 0$ limit
\beq
S &\rightarrow& a^4 \sum_x \sum_{M < N} \frac{\beta g_5^2}{2N_c}  \mathrm{Tr} F_{M N}^2 (x) + O(a^6).
\eeq
Therefore, one choice of $\beta$ to reproduce the action in the continuum limit can be given by $\beta=2N_c a/g_5^2$ at tree level.
Note that the lattice spacing ($a$) in the numerator is added in contrast with four-dimensional case.

The lattice spacing is just a regulator which is introduced by hand, so that there is no reason to have a five-dimensional isotropy.
In general, we can introduce a different lattice spacing for the fifth dimension ($a_5$) from the one of the other dimensions ($a_4$), namely $a_4 \ne a_5$ is allowed.
The corresponding lattice gauge action is also modified by two independent normalization factors ($\beta_4$ and $\beta_5$),
\beq
S&=& \beta_4 \sum_{x} \sum_{1 \le \mu \le \nu \le 4} \left( 1-\frac{1}{N_c} \mathrm{Tr} U_{\mu \nu} \right)
+ \beta_5 \sum_{x} \sum_{1 \le \mu \le 4} \left( 1-\frac{1}{N_c} \mathrm{Tr} U_{\mu 5} \right).\label{eq:aniso-action}
\eeq
Here the link variables for the four-dimensional components and the fifth direction are given by
\beq
U_\mu (x)&=& e^{i a_4 g_5 A_\mu (x+\hat{\mu}/2)}, \mbox{ for $\mu=1,\cdots, 4$},
\nonumber\\
U_5 (x) &=&e^{i a_5 g_5 A_5 (x+\hat{5}/2)} ,
\eeq
respectively.
The plaquettes in four-dimensional subspaces and in $\mu$-$5$ plane are defined as follows:
\beq
U_{\mu \nu}(x) &=&U_{\mu}(x) U_{\nu}(x+\hat{\mu}a_4) U^\dag_\mu (x+\hat{\nu}a_4)U^\dag_\nu(x),\nonumber\\
U_{\mu 5}(x)    &=&U_{\mu}(x) U_{5}(x+\hat{\mu}a_4) U^\dag_\mu (x+\hat{5}a_5)U^\dag_5(x).\label{eq:aniso-link}
\eeq
Substituting Eqs. (\ref{eq:aniso-link}) into Eq. (\ref{eq:aniso-action}), we find that one possible naive continuum limit ($a_4,a_5 \rightarrow 0$)
is obtained by choosing
\beq
\beta_4 =2 N_c \frac{a_5}{g_5^2}, ~~~~
\beta_5 =2 N_c \frac{a_4}{g_5^2} \frac{a_4}{a_5}.
\eeq
In the practical lattice numerical simulation, we utilize the lattice bare coupling constant ($\beta$), which is normalized a four-dimensional UV cutoff, and the anisotropy
parameter ($\gamma$),
\beq
\beta=2N_c \frac{a_4}{g_5^2},~~~~
\gamma=\frac{a_4}{a_5}\label{eq:def-gamma}.
\eeq
Finally, we obtain the anisotropic plaquette gauge action,
\beq
S&=& \frac{\beta}{\gamma} \sum_{x} \sum_{1 \le \mu \le \nu \le 4} \left( 1-\frac{1}{N_c} \mathrm{Tr} U_{\mu \nu} \right)
+ \beta \gamma \sum_{x} \sum_{1 \le \mu \le 4} \left( 1-\frac{1}{N_c} \mathrm{Tr} U_{\mu 5} \right).
\eeq

In this work, firstly we investigate the phase structure for the isotropic plaquette gauge action (Eq.~(\ref{eq:iso-action})) in Sec.~\ref{sec:iso-lattice}, and then study the anisotropic case (Eq.~(\ref{eq:aniso-action})) in Sec.~\ref{sec:aniso-lattice}.
The large anisotropy parameter region ($\gamma > 1 $) describes the left geometrical picture in Fig.~\ref{fig:image} when the lattice extent in the fifth dimension is not larger than the one in the other dimensions.
On the other hand, we expect that the right picture in Fig.~\ref{fig:image} would be induced in a small $\gamma$ region.

\subsection{Simulation setup and observables}
Gauge configurations are generated by the pseudo-heatbath algorithm with the
over-relaxation, mixed in the ratio of~$1:5$. We call one pseudo-heatbath
update sweep plus five over-relaxation sweeps as a ``Sweep''.
The number of Sweeps for the measurements is~$5,000$ after the thermalization.
Statistical errors are estimated by the jackknife method.

We generate the configurations starting with ``cold start", which is in the ordered phase and we set the initial configuration unity. 
We also utilize ``hot start" which is in the disordered phase and whose initial configuration is random.
It is known that the numerical simulation results using the four-dimensional plaquette gauge action are independent of the initial configuration after the thermalization.
However, a hysteresis is observed in the five-dimensional SU($2$) plaquette gauge action~\cite{Creutz:1979dw}.
The hysteresis comes from lattice artifacts.
Our work gives the first numerical simulation results for the five-dimensional SU($3$) plaquette gauge action.
We firstly investigate the existence of the phase transition and the hysteresis.

We measure four quantities, namely plaquettes ($P_{ss}$ and $P_{st}$) which are defined in the four-dimensional Euclidean space-time and the $\mu$--$5$ plane with $\mu=1, \cdots , 4$, and the Polyakov loops (Ploop$_s$ and Ploop$_t$) in the fourth and the fifth directions.
\beq
\langle P_{ss} \rangle = \langle \frac{1}{6 N_c V_5} \sum_{x} \sum_{\mu,\nu \ne 5} U_{\mu \nu} (x) \rangle,&&
\langle P_{st} \rangle = \langle \frac{1}{4 N_c V_5} \sum_{x} \sum_{\mu } U_{\mu 5} (x) \rangle, \\
\langle \mathrm{Ploop}_{s}  \rangle = \langle \frac{N_s}{N_c V_5} \sum_{\vec{x},x_5} \prod_{x_4=1}^{N_s} U_{4} (\vec{x},x_4, x_5) \rangle, &&
\langle \mathrm{Ploop}_{t}  \rangle = \langle \frac{N_t}{N_c V_5} \sum_{\vec{x},x_4} \prod_{x_5=1}^{N_t} U_{5} (\vec{x},x_4, x_5) \rangle.\nonumber\\ \label{eq:def-Ploopt}
\eeq
Here $V_5$ denotes the five-dimensional lattice volume ($V_5=(N_s)^4 \times N_t$), and $N_s$ and $N_t$ denote the lattice extent for the isotropic four dimensions and the independent fifth dimension, respectively.
The variables $x$ and $\vec{x}$ denote a five-dimensional coordinate ($x=(x_1,x_2,x_3,x_4,x_5)$) and a three-dimensional spatial coordinate ($\vec{x}=(x_1,x_2,x_3)$), respectively.

In this theory there are two global symmetries, namely the $Z_3$ center symmetry and the global SU($3$) gauge symmetry.
The order parameter of the center symmetry is given by the magnitude of the Polyakov loop.
If the magnitude of the Polyakov loop is zero, then the center symmetry is preserved.
The phase corresponds to the ``confined phase", since the magnitude of the Polyakov loop is related to the free energy to take a single quark infinity.
We expect that it occurs in the strong coupling limit ($\beta \rightarrow 0$), since the strong expansion can describe the confinement.
On the other hand, if the quantity is not zero, then the center symmetry is spontaneously broken.
This phase is called ``deconfined phase".
At the tree level, the vacuum configuration in the deconfined phase can be given by the unity matrix ($U_M=\mathbb{I}$).
In the quenched SU($3$) gauge theory, because of the $Z_3$ center symmetry the following Polyakov loops,
\beq
\langle \mathrm{Ploop} \rangle = e^{2 \pi i l/3},
\eeq
for $l=0,1,2$ are equivalent.

\begin{figure}[h]
\vspace{0.5cm}
\begin{center}
\includegraphics[scale=0.35]{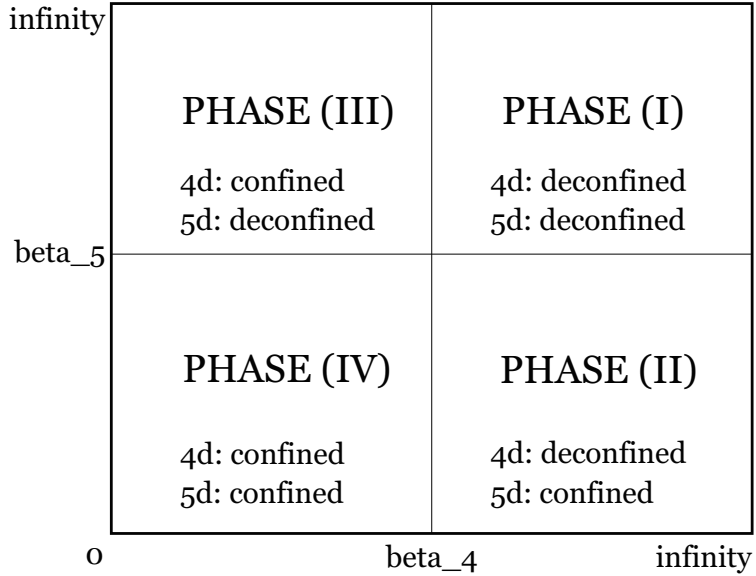}
\caption{Conjectured phase diagram for the anisotropic five-dimensional lattice gauge theory}
\label{fig:conj-phase-diagram}
\end{center}
\end{figure}
Figure~\ref{fig:conj-phase-diagram} shows a naive expectation for the phase structure.
If we introduce the anisotropy, there are two order parameters for the confined/deconfined phase transition in both four-dimensional spaces and fifth dimension.
We expect that there must be the deconfined phase in the weak coupling region, while the quarks and the gluons are confined in the strong coupling regime for both order parameters.
Correspondingly the PHASE (I) (right-top phase in Fig.~\ref{fig:conj-phase-diagram}) can be called ``deconfined phase", and PHASE (IV) (left-bottom phase) is ``confined phase"  as a whole five-dimensional QCD.
The PHASE (II) (right-bottom phase) and PHASE (III) (left-top phase) will not appear in the isotropic lattices.
In PHASE (II) the quark currents are not confined on the four-dimensional brane, while they cannot extend to the fifth direction.
Thus the QCD dynamics is localized on the four-dimensional brane.
We expect that PHASE (II) can be realized in the small $\gamma$ regime, where $a_5 > a_4$.
On the other hand, in PHASE (III) the confinement occurs in four dimensions while the quarks are deconfined in the fifth direction.
We expect that the phase should appear when the physical length of the fifth dimension is smaller than the other four-dimensional spaces.
The geometry corresponds to the $4+1$ dimensional spaces, where one dimension is compact space.
The KK modes of the fifth component of the gauge field can be interpreted as an adjoint scalar field in four dimensions.
Therefore the theory would give the same phase structure with the one of the SU($3$) gauge theory coupled to the adjoint scalar fields~\cite{Fradkin:1978dv}.

Another order parameter is related with the global SU(3) gauge symmetry.
It is an important aspect to understand the spontaneous gauge symmetry breaking, in particular, within the gauge-Higgs unification or the grand unified theory based on the extra dimension model.
In $D+1$ gauge theory where one dimension is compact space, if the Wilson line phase (non-abelian Aharonov-Bohm (AB) phase) in the compact dimension have a nontrivial eigenvalue, then the gauge symmetry in $D$ dimensions can be spontaneously broken via the Hosotani mechanism~\cite{Hosotani:1983xw}.
It is known that in the SU($3$) gauge theory coupled to adjoint fermions with the periodic boundary condition in $3+1$ dimensions, there are two possible phases with the different global gauge symmetry in the deconfined phase~\cite{Cossu:2009sq,Cossu:2013ora, Cossu:2013nla}. 
However, according to the discussion based on the perturbative one-loop effective potential, it is known that the pure gauge theory in any $D+1$ dimensions never shows the spontaneous gauge symmetry breaking~\cite{Gross:1980br, Weiss:1980rj, Hosotani:1983xw, Kashiwa:2013rmg}.
In fact, we do not find the spontaneous global gauge symmetry breaking in our nonperturbative results.

\section{Phase structure of the isotropic SU($3$) lattice gauge theory in five dimensions}\label{sec:iso-lattice}
Let us start with the isotropic lattices.
The numerical simulation of the five-dimensional SU($2$) plaquette gauge action shows the bulk first order phase transition in the pioneer work by M.~Creutz~\cite{Creutz:1979dw}.
The expectation value of plaquette shows the hysteresis at $\beta \approx 1.65$ in the five-dimensional SU($2$) lattice gauge theory, while it smoothly connects from the weak coupling region to the strong coupling region in the four-dimensional case.
In this section, we carry out the numerical simulation utilizing the five-dimensional SU($3$) plaquette gauge action on the isotropic lattice ($N_s=N_t$ and $\gamma=1.00$).
We investigate $1.0 \le \beta \le 9.5$ region.

\begin{figure}[h]
\begin{center}
\includegraphics[scale=0.45]{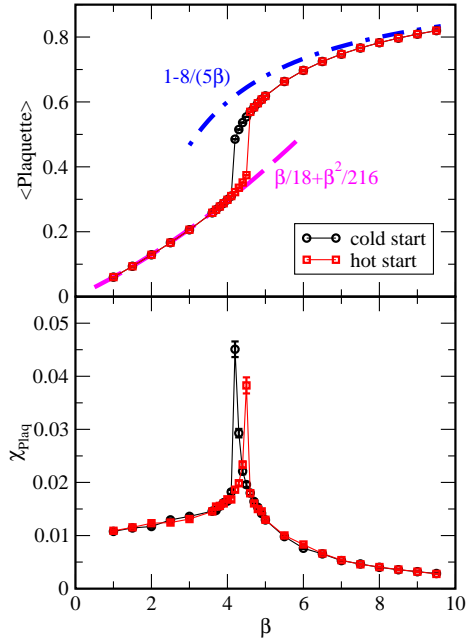}
\caption{
Vacuum expectation values of the plaquette and its susceptibility as a function of $\beta$
 for $N_s=N_t=8$ and $\gamma=1.00$.
The cycle (black) symbol denotes the results for starting with ``cold start", while the square (red) symbol denotes the ones for starting with ``hot start".
The dot-dashed (blue) and dashed (magenta) lines on the top panel show the results of the weak (Eq.~(\ref{eq:weak-coupling})) and strong coupling expansions (Eq.~(\ref{eq:strong-coupling})), respectively. }
\label{fig:L-8-T-8-G1.0-2}
\end{center}
\end{figure}
Figure~\ref{fig:L-8-T-8-G1.0-2} and \ref{fig:L-8-T-8-G1.0-3} show the simulation results for the $N_s=8$ lattice extent. We also give raw data in Appendix~\ref{sec:raw-data}.
The top (bottom) panel in Fig.~\ref{fig:L-8-T-8-G1.0-2} shows the expectation value of the plaquette (its susceptibility) as a function of $\beta$.
Figure~\ref{fig:L-8-T-8-G1.0-3} shows the magnitude of the Polyakov loop in the same range of $\beta$.
The cycle (black) symbol denotes the data of starting with ``cold start". The corresponding configuration lives in the ordered phase, and we set all initial link variables to unity.
On the other hand, the square (red) symbol denotes the ones starting with ``hot start". The corresponding configuration is in the disordered phase, and the initial link variable is a random number.
The dot-dashed (blue) and dashed (magenta) lines on the top panel in Fig.~\ref{fig:L-8-T-8-G1.0-2} show the results of the weak and strong coupling expansions given by the following equations;
\beq
\mbox{weak coupling: } && \langle \mathrm{plaquette} \rangle = 1-\frac{8}{5 \beta},\label{eq:weak-coupling} \\
\mbox{strong coupling: } && \langle \mathrm{plaquette} \rangle =  \frac{\beta}{18}+ \frac{\beta^2}{216}, \label{eq:strong-coupling}
\eeq
for the five-dimensional SU($3$) plaquette gauge theory.

\begin{figure}[h]
\begin{center}
\includegraphics[scale=0.45]{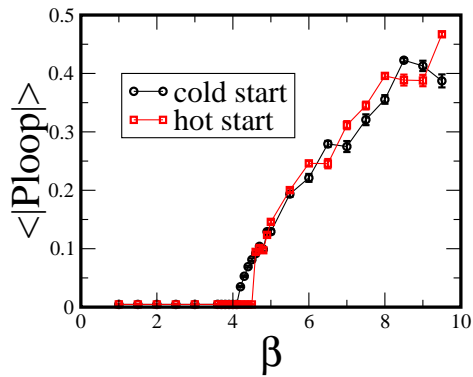}
\caption{
Vacuum expectation values of the magnitude of
 the Polyakov loop as a function of $\beta$
 for $N_s=N_t=8$ and $\gamma=1.00$.
The cycle (black) symbol denotes the results for starting with ``cold start", while the square (red) symbol denotes the ones for starting with ``hot start".
 }
\label{fig:L-8-T-8-G1.0-3}
\end{center}
\end{figure}
We found that there is a hysteresis in the range of $4.2 \le \beta \le 4.5$ in the SU($3$) plaquette gauge action in all panels.
Furthermore, the susceptibility of the plaquette value shows a sharp peak at $\beta=4.2$ for cold start and $\beta=4.5$ for hot start.
Clear hysteresis is an evidence of a first order phase transition.

In the region of $\beta$ lower than the regime where the hysteresis exists, the average of the plaquette is consistent with the result of the strong coupling expansion, and the magnitude of the Polyakov loop is approximately zero.
Therefore the region corresponds to a confined phase, and the center symmetry is preserved there.
On the other hand, in the region of $\beta$ higher than the regime where the hysteresis exists, the plaquette value approaches the result of the weak coupling expansion.
The magnitude of the Polyakov loop becomes nonzero values, therefore we can conclude
the center symmetry is spontaneously broken.
At even higher $\beta$ ($\beta > 9.5$) or lower $\beta$ ($\beta < 1.0$), the plaquette values are already consistent with the analytical results, so that we expect that there is no more phase transition.
From now, we focus on the region where the hysteresis is observed.

\begin{figure}[h]
\vspace{0.5cm}
\begin{center}
\includegraphics[scale=0.4]{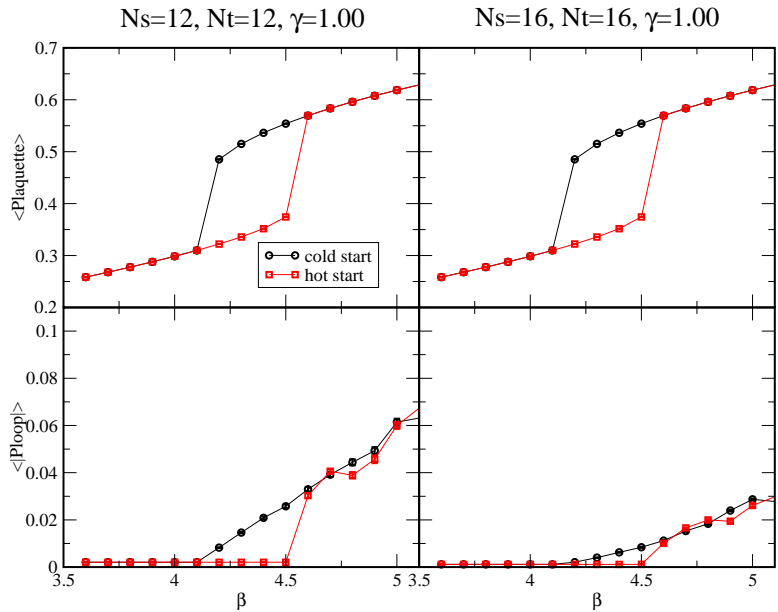}
\caption{Vacuum expectation values of the plaquette (top panels) and the magnitude of the Polyakov loop (bottom panels) as a function of $\beta$ for the isotropic lattices with $N_s=12$ (left panels) and $16$ (right panels). The cycle (black) symbol denotes the results for starting with ``cold start", while the square (red) symbol denotes the ones for starting with ``hot start".
}
\label{fig:L-12-L-16-G1.0}
\end{center}
\end{figure}

To clarify the origin of the first order phase transition,
we change the lattice volume and investigate the volume dependence of the critical value of $\beta$ ($\beta_c$) where the hysteresis appears.
Figure~\ref{fig:L-12-L-16-G1.0} shows the expectation values of the plaquette and the magnitude of the Polyakov loop for $N_s=12$ and $16$.
If the phase transition is a physical (thermal) one, then the critical value of $\beta$ should depend on the volume.
However, there is no volume dependence of $\beta_c$ in this first order phase transition.
We can conclude that this is the bulk first order phase transition coming from a lattice artifact\footnote{However, the magnitude of the Polyakov loop at the fixed $\beta$ becomes small in the large volume. It might suggest that a physical phase transition is hidden in the range where the bulk phase transition occurs. To find it, we need the larger lattice simulation as a future work.}.

In order to avoid the bulk phase transition, there are at least two possible ways to study on the five-dimensional gauge theories.
The first one is to change the lattice gauge action.
Generally, the existence of the bulk phase transition depends on the lattice gauge action.
The different lattice gauge action which gives the same continuum limit might solve this problem~\cite{Kawai:1992um}.
Another way to avoid the bulk phase transition is the introduction of the anisotropy into the lattice spacings~\cite{Ejiri:2000fc}. 
There are several works on the anisotropic pure SU($2$) lattice gauge theory \cite{Ejiri:2000fc}--\cite{Dimopoulos:2001un}.
The region, where the hysteresis disappears, exists in both large and small $\gamma$ regions.
In the next section, we also introduce the anisotropy, and determine the critical values of the parameters which give boundaries of the phases in the SU($3$) plaquette gauge action.

\section{Phase structure of the anisotropic SU($3$) lattice gauge theory in five dimensions}\label{sec:aniso-lattice}
Now, we introduce the anisotropy between the four-dimensional spaces and the fifth dimension on the lattice.
Firstly, we introduce the anisotropy parameter ($\gamma$), which gives the ratio of the fourth- and the fifth-dimensional lattice spacings at the tree level (Eq.(\ref{eq:def-gamma})), with keeping the isotropic lattice extent ($N_s=N_t$).
Next, we investigate the phase structure for the anisotropic lattice extent ($N_s > N_t$) with the anisotropic lattice spacing.
We found that the phase structure concerning the center symmetry for the anisotropic lattice extent is qualitatively similar with the one for the isotropic lattice extent.
Thus there are four phases corresponding to Fig.~\ref{fig:conj-phase-diagram}.
We also investigate $N_s$ and $N_t$ dependences of the phase structure.
In particular, we determine the critical value of $\beta$ where the center symmetry is the spontaneously broken for each value of $\gamma$.
Furthermore, we also determine two critical values of $\gamma$, namely $\gamma_c^{(l)}$ and $\gamma_c^{(s)}$.
In the range of $\gamma > \gamma_c^{(l)}$ and $\gamma< \gamma_c^{(s)}$, the values of $\beta_c$ for the Polyakov loops in the fourth- and the fifth-direction show a discrepancy in large and small $\gamma$ regions.
The strong hysteresis also disappears in these regions.

\subsection{Phase structure of the isotropic lattice extent with the anisotropic lattice spacings}
Let us introduce the anisotropy parameter ($\gamma$) on the isotropic lattice extent ($N_s=N_t=8$).
We carry out the simulations for more than $15$ values of $\gamma$ in $0.40 \le \gamma \le 4.00$ region.
We observe the values of plaquettes ($P_{ss}$ and $P_{st}$) and the Polyakov loops (Ploop$_s$ and Ploop$_t$).
The properties of plaquettes are very similar with the ones
of the Polyakov loops, so that we show only the data of the Polyakov loops
from here.

\begin{figure}[h]
\begin{center}
\includegraphics[scale=0.4]{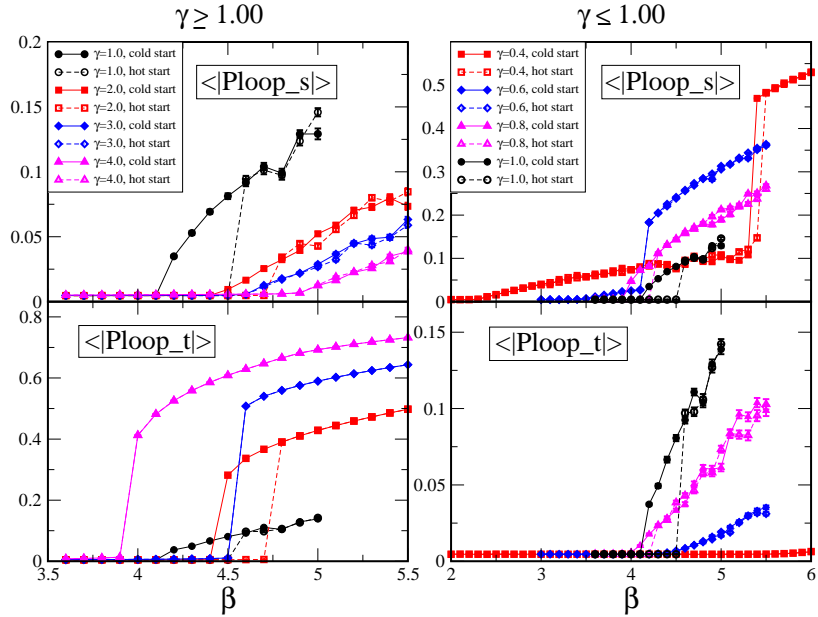}
\caption{Vacuum expectation values of the magnitude of the Polyakov loop for each direction on the $N_s=N_t=8$ lattice. Left (right) panels show the results for $\gamma \ge 1.00$ ($\gamma \le 1.00$), and top (bottom) panels describe $\langle |$Ploop$_s | \rangle$ ($\langle |$Ploop$_t | \rangle$). }
\label{fig:raw-data-Ploop-L-8-T-8-gamma}
\end{center}
\end{figure}

We determine the critical value of $\beta$ ($\beta_c$) for the spontaneous center symmetry breaking from the magnitude of Polyakov loops for each value of $\gamma$.
Figure~\ref{fig:raw-data-Ploop-L-8-T-8-gamma} shows that the examples of the magnitude of Ploop$_s$ and Ploop$_t$ as a function of $\beta$.
Firstly, we found that the Polyakov loop in the fifth direction is a drastically changing at $\gamma \ge 1.00$ rather than $\gamma \le 1.00$, while the one in the fourth direction indicates a strong phase transition in $\gamma \le 1.00$ rather than $\gamma \ge 1.00$.
Secondly, there are two transition points for the Polyakov loop in fourth direction ($\langle | \mathrm{Ploop}_s |\rangle$) in $\gamma=0.40$ and $0.60$; {\it e.g.} $\beta=2.40$ and $5.40$ in $\gamma=0.40$ and $\beta=3.60$ and $4.20$ in $\gamma=0.60$, respectively.
Beyond the former critical $\beta$, the value of  $\langle |$Ploop$_s| \rangle$ smoothly changes from zero to nonzero, and at the later critical point it shows the gap.
Furthermore, the later transition coincides with the phase transition for $\langle |$Ploop$_t | \rangle$.
We also found that if the value of $\gamma$ is around unity, still there are hysteresis in all panels.
On the other hand, the data with $\gamma = 3.00, 4.00, 0.60$ and $0.40$ do not show the strong hysteresis.
It suggests that the order of the phase transition would be changed to the second order or a weakly first phase transition from the bulk first order one.

\begin{figure}[h]
\begin{center}
\includegraphics[scale=0.5]{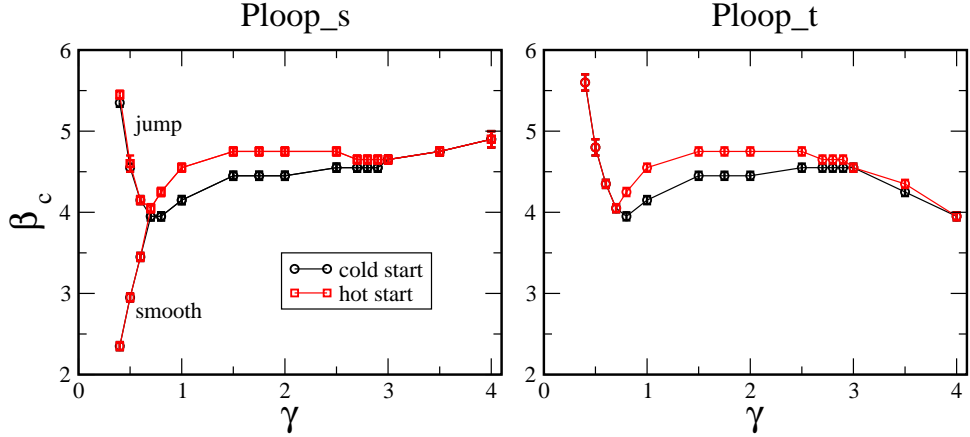}
\caption{Values of $\beta_c$ which are determined by the expectation values of Polyakov loop (Ploop$_s$ and Ploop$_t$) on the $N_s=N_t=8$ lattice as a function of $\gamma$. Error bar is estimated by the interval of the parameters we measured. There are two lines for the Polyakov loop in fourth direction (Ploop$_s$)  in $\gamma < 0.70$ region. At the lower $\beta_c$, there is a smooth changing of the Polyakov loop, while a jump is observed at the higher one. }
\label{fig:L-8-T-8-betac}
\end{center}
\end{figure}
Figure~\ref{fig:L-8-T-8-betac} shows the summary of $\beta_c$ for Ploop$_s$ and Ploop$_t$ as a function of $\gamma$.
The magnitude of the Polyakov loop is approximately zero below each line, while that is nonzero above the line.
Namely, the high $\beta$ region corresponds to the center broken (deconfined) phase, while the low $\beta$ region is in the center symmetric (confined) phase.
We found that the values of $\gamma_c^{(l)}$ is $2.90 \pm 0.10$ and $\gamma_c^{(s)}=0.70 \pm 0.10$ for $N_s=N_t=8$, where the strong hysteresis disappears.
Here we estimate the error size from the interval of the parameter we measured.

\begin{figure}[h]
\begin{center}
\includegraphics[scale=0.3]{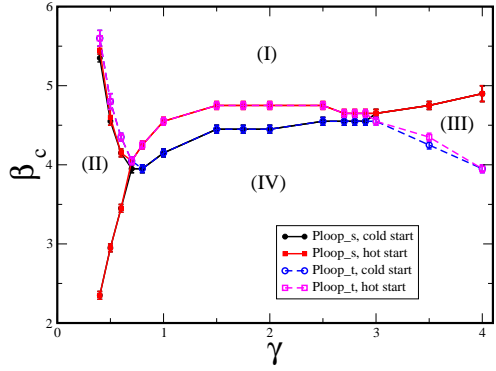}
\caption{Phase diagram for $N_s=N_t=8$ lattice. The notation (I) -- (IV) corresponds to the PHASE (I) -- (IV) in Fig.~\ref{fig:conj-phase-diagram}. }
\label{fig:phase-L-8-T-8}
\end{center}
\end{figure}
We obtain the phase diagram in Fig.~\ref{fig:phase-L-8-T-8} by overlapping two panels in Fig.~\ref{fig:L-8-T-8-betac}.
We clearly see that there are four phases corresponding to Fig.~\ref{fig:conj-phase-diagram}.
In the small $\gamma$ region, there are three phases, namely both the four-dimensional and the fifth dimensional quark currents are confined (PHASE (IV)) and deconfined (PHASE (I)), and only the quark current in the fifth 
direction is confined while the confinement does not occur in four-dimensional space (PHASE (II)).
On the other hand, in the large $\gamma$ region, there are also three phases, namely the five-dimensional confined (PHASE (IV)) and deconfined (PHASE (I)) phases, and only in four-dimensional spaces
the confinement occurs while there remains a dynamics of quarks in the fifth direction (PHASE (III)).
Note that there is no direct phase transition between PHASE~(II) and PHASE~(III) in the whole region.

\subsection{Phase structure of the anisotropic lattice extent}\label{sec:L8-T4}
Now, let us consider the anisotropic lattice extent, namely $N_s > N_t$.
One of the aims of introducing the anisotropic lattice extent is to use an analogy with the finite temperature system in $3 + 1$ dimensions.
Here the short lattice extent is interpreted as a compactification radius of the extra dimension instead of the temperature.
Furthermore, it is known that the region, where the bulk phase transition appears, shrinks in the five-dimensional SU($2$) gauge theory~\cite{Ejiri:2000fc}, if the anisotropic lattice extent is introduced.
The other practical advantage of introducing the anisotropic lattice extent is that we can reduce the simulation cost if the qualitative phase diagram is the same with the isotropic lattice extent.
In this section, firstly we study on the phase structure for $N_s=8$ and $N_t=4$ with the anisotropy parameter, and compare the results with the previous section.
Furthermore, we show the $N_s$ and $N_t$ dependences of the critical values of the parameters.
Finally, we obtain the whole phase diagram for the SU($3$) plaquette gauge action on $\beta_4$ -- $\beta_5$ plane.

\begin{figure}[h]
\begin{center}
\includegraphics[scale=0.3]{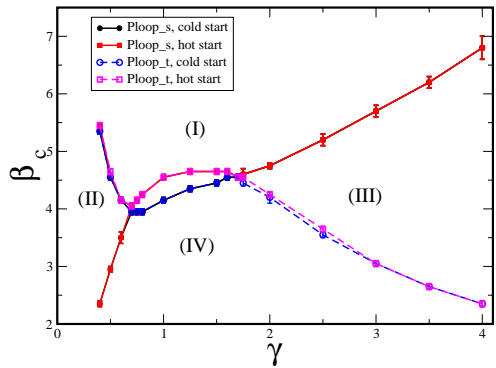}
\caption{Values of $\beta_c$ as a function of $\gamma$ determined by the magnitudes of Ploop$_s$ and Ploop$_t$ for $N_s=8, N_t=4$ lattice. The notation (I) -- (IV) corresponds to the PHASE (I) -- (IV) in Fig.~\ref{fig:conj-phase-diagram}. }
\label{fig:phase-L-8-T-4-gamma}
\end{center}
\end{figure}
Figure~\ref{fig:phase-L-8-T-4-gamma} shows the critical values of $\beta$ for
$N_s=8,N_t=4$ as a function of $\gamma$.
As with the results for the isotropic lattice extent, there are four phases in total depending on $\gamma$ and $\beta$.
The differences from the result of Fig.~\ref{fig:phase-L-8-T-8} are the value of $\gamma_c^{(l)}$ where the hysteresis disappears in the large $\gamma$ region and the values of $\beta_c$ in $\gamma > \gamma_c^{(l)}$ regime.
In the case of $N_t=4$, the critical $\gamma$ is around $\gamma_c^{(l)} = 1.70^{+0.05}_{-0.10}$, while $\gamma_c^{(l)}=2.90 \pm 0.10$ in the case of $N_t=8$.
On the other hand, the smaller critical $\gamma$ ($\gamma_c^{(s)}$) is the same in both cases, namely $\gamma_c^{(s)}=0.70^{+0.05}_{-0.10}$.

We also investigate the volume dependence.
\begin{table}[t]
\begin{center}
\begin{tabular}{|c||c|c||c|c|}
\hline
$(N_s,N_t)$ & $\gamma_c^{(l)}$ &  $\beta_c$ at $\gamma_c^{(l)}$ & $\gamma_c^{(s)}$ & $\beta_c$ at $\gamma_c^{(s)}$ \\
\hline
$(8,8)$    &  $2.90 \pm 0.10$                 &   $4.60 \pm 0.10$  & $0.70 \pm 0.10$ &  $4.00 \pm 0.10$ \\ 
$(8,4)$    &  $1.70 ^{+ 0.05}_{-0.10} $ &   $4.55 \pm 0.05$  & $0.70 \pm 0.10$ &  $4.00 \pm 0.10$ \\ 
$(12,4)$    &  $1.70 ^{+0.05}_{-0.10}$ &    $4.55 ^{+ 0.15}_{-0.05}$  & $0.70 ^{+0.05}_{-0.10}$ &  $4.00 \pm 0.15$ \\ 
$(16,4)$    &  $1.50 ^{+ 0.20}_{-0.25}$ &    $4.65 ^{+0.05}_{-0.25}$  & $0.75 ^{+0.05}_{-0.15}$ &  $4.00 ^{+0.20}_{-0.10}$ \\ 
$(12,6)$    &  $2.30 ^{+0.20}_{- 0.10}$ &    $4.65 ^{+ 0.05}_{-0.15}$  & $0.70 \pm 0.10$ &  $4.00 \pm 0.10$ \\ 
\hline
\end{tabular}
\caption{Values of $\gamma_c^{(l)}$, $\gamma_c^{(s)}$ and $\beta_c$ at the critical $\gamma$ for each lattice size. Error size is estimated by the interval of the parameters we measured.}
\label{table:gamma_c} 
\end{center}
\end{table}
The phase diagrams for $(N_s,N_t)=(12,4),(16,4)$ and $(12,6)$ are given in Appendix~\ref{sec:app-phase-diagram}.
Moreover, the values of $\gamma_c^{(l)}$, $\gamma_c^{(s)}$ and $\beta_c$ at each $\gamma_c$ are summarized in Table.~\ref{table:gamma_c}.

In the large $\gamma$ region, although the critical values of $\beta$ of the phase transition between PHASE (I) and (III) depend on both $N_s$ and $N_t$, the one between
PHASE (III) and (IV), which describes the phase transition for the center symmetry in the fifth dimension, depends on only $N_t$.
The value of $\gamma_c^{(l)}$ strongly depends on $N_t$, but is independent of $N_s$ within errorbar.
On the analogy of the finite temperature phase transition in $3+1$ dimensional lattice simulation, the $N_t$ dependence of $\gamma_c^{(l)}$ suggests that a ``thermal" phase transition appears in this region~\cite{Ejiri:2000fc, deForcrand:2010be,DelDebbio:2012mr}.

On the other hand, in the small $\gamma$ limit, namely $a_5$ is much larger than $a_4$, we expect that the four-dimensional brane decouples with each other.
The number of four-dimensional brane does not give any effects, so that $\gamma_c^{(s)}$ is independent of $N_t$.
In $\gamma \le \gamma_c^{(s)}$ region, only PHASE (I), (II) and (IV) appear.
The phase transition between PHASE (I) and (II) depends on only $N_t$, while the transition from PHASE (II) to (IV) depends on only $N_s$.
Furthermore the values of $\gamma_c^{(s)}$ and $\beta_c$ at the point are independent of both $N_s$ and $N_t$ as shown in Table~\ref{table:gamma_c}.
The property would suggest that the critical point depends only on the ratio of the lattice spacing, namely the hierarchy of UV cutoff between fourth and fifth dimensions.
We expect that the layered structure is valid in PHASE (II)~\cite{Dimopoulos:2001un, Nicolis:2010nh, Knechtli:2011gq}

Finally, we obtain the phase diagram for the five-dimensional pure SU($3$) lattice gauge theory on $\beta_4$ -- $\beta_5$ plane.
\begin{figure}[h]
\begin{center}
\includegraphics[scale=0.5]{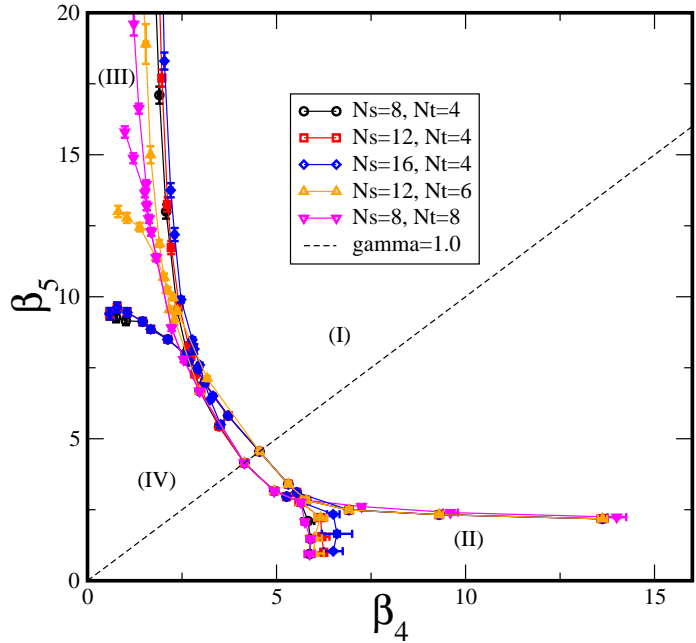}
\caption{Phase structure for SU($3$) anisotropic lattice gauge theory on $\beta_4$ -- $\beta_5$ plane. The notation (I) -- (IV) corresponds to the PHASE (I) -- (IV) in Fig.~\ref{fig:conj-phase-diagram}. }
\label{fig:phase-beta4-beta5}
\end{center}
\end{figure}
On the diagonal line ($\beta_4=\beta_5$), there are only PHASE (I) and (IV), and the phase transition is the bulk one.
However, when the anisotropy becomes strong, new two phases appear.
With the relatively small fifth lattice extent on the lattice, where the $4+1$ dimensional geometry with one compact dimension consists, PHASE (III) shows up.
To study the phenomenological gauge-Higgs unification or universal extra dimension models is possible in this geometrical lattice setup.
The opposite direction, where $\beta_5$ is small, has PHASE (II).
In this phase, we expect that the gauge and quark fields are localized on the four-dimensional layered brane because of the nonperturbative effect.
The contributions coming from the other branes can be negligible.
The critical point, where the bulk first order phase transition ends, would change the order of the transition to the second order phase transition.
The existence of the second order critical end point even for the SU($2$) gauge theory is still under investigation~\cite{DelDebbio:2013rka}, and we need the large lattice data to show it.

\section{Appearance of the center domains}\label{sec:meta-stable}
During the numerical simulation, we found that strange configurations sometimes appear.
The complex phase of the Polyakov loop in fifth direction of these configurations 
is located at $\pm \pi/3$ or $\pi$.
In the previous section, we drop such configurations from our analysis.

At first, we considered it was ``split  (skewed) phase", which is known in other theories, {\it e.g.} the two-dimensional Wilson line model as one of the dimensionally reduced models of the three-dimensional SU($3$) gauge theory~\cite{Bialas:2004gx} and the SU($3$) gauge theory coupled to the adjoint fermions with the periodic boundary condition in $3+1$ dimensions~\cite{Cossu:2009sq}.
However, it is happened by the different reason~\footnote{P.~de~Forcrand and O.~Akerlund told us the point after we had submitted this draft on arXiv.}, namely the coexistence of the Polyakov loops with different $Z_3$ charges of the center symmetry.
Such configurations are discussed in the finite temperature QCD to explain the small shear viscosity and the large opacity of the Quark-Gluon-Plasma (QGP) in the experiment~\cite{Asakawa:2012yv}, and
the lattice simulations for both the quenched and full QCD in $3+1$ dimensions are also investigated~\cite{Borsanyi:2010cw,Endrodi:2014yaa}.
We show the detailed results for the configurations with more than one center domain.

\begin{figure}[h]
\begin{center}
\includegraphics[scale=0.4]{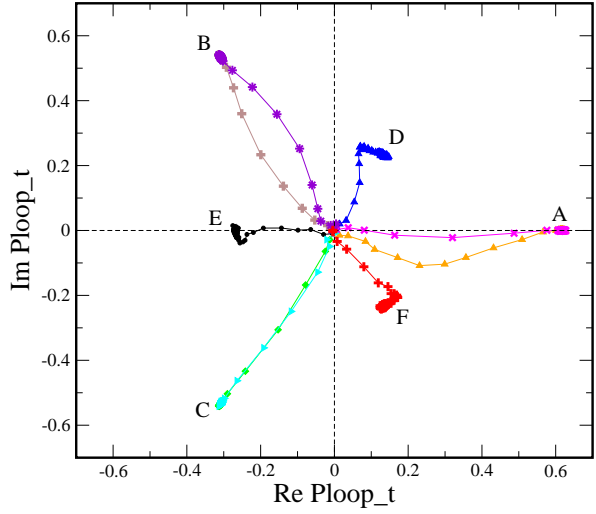}
\caption{ Histories of the Polyakov loop in $t$ direction for $N_s=8, N_t=4, \beta=3.50$ and $\gamma=3.00$. The different symbol (and color) describes the independent configurations starting with the different random numbers. Each point denotes the value of Polyakov loop for each Sweep. Symbols ``A--F" denote the labels of the location of the vacuum after the thermalization.}
\label{fig:dis-Ploopt}
\end{center}
\end{figure}
Figure~\ref{fig:dis-Ploopt} shows the examples of the histories of the
Polyakov loop distributions for the first $500$ Sweeps in our numerical simulation.
The lattice setup is ($N_s,N_t,\beta, \gamma$)$=$ ($8,4,3.50,3.00$).
Each symbol (color) shows the independent configuration starting with the different random number.
The distribution of the Polyakov loop starts around the origin in the complex plane since the initial configuration is just a random number for each lattice site.
We carry out the simulations starting with more than $200$ independent random numbers.
After first few number of Sweeps, most of the configurations move to the usual $Z_3$ symmetric phase whose complex phase is $0$ or $\pm 2\pi/3$ (A, B and C in Fig.~\ref{fig:dis-Ploopt}).
However, we also found some configurations, whose appearance probability is roughly $4\%$, move to the vacuum where the complex phase of the Polyakov loop is $\pm \pi/3$ or $\pi$ (D, E and F in the same figure).
The magnitude of Polyakov loop of the configurations in the vacua D,E and F is roughly $1/3$ times in the contrast of the ones in the vacua A,B and C with the same value of $\beta$.
These properties are consistent with the ones of the split phase in the theory coupling to the adjoint fermion in $3+1$ dimensions~\cite{Cossu:2009sq}.

To study in detail, we also show the average of the Polyakov loop over only one spatial slice at the fixed $x_\mu (\mu=1,\cdots, 4)$ for each spatial direction,
\beq
 \mathrm{Ploop}_{t} (x_\mu)  &=&  \frac{N_s N_t}{N_c V_5} \sum_{x_\nu (\nu \ne \mu)} \prod_{x_5=1}^{N_t} U_{5} (x_\nu, x_5) ,\label{eq:Ploopt-each-slice} \\
 &=& | \mathrm{Ploop}_{t}| (x_\mu) e^{i\theta (x_\mu)}\label{eq:def-theta}.
\eeq
We found that the phase $\theta (x_\mu)$ for one $\mu$-direction strongly depends on the site for the configurations in the vacua D, E and F, although the magnitude $| \mathrm{Ploop}_{t}|(x_\mu)$ does not depend on the site so much.
\begin{figure}[h]
\begin{center}
\includegraphics[scale=0.3]{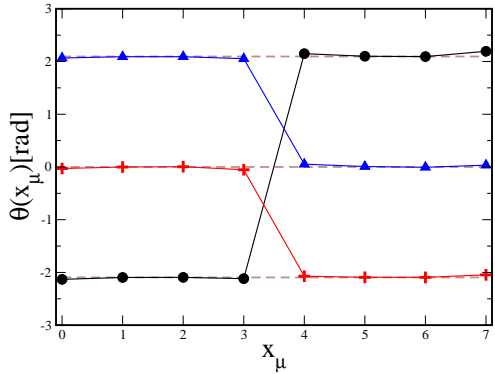}
\caption{ Phase $\theta(x_\mu)$ of the averaged Polyakov loop over only one spatial slice (Eq.~(\ref{eq:Ploopt-each-slice})) for $N_s=8, N_t=4, \beta=3.50$ and $\gamma=3.00$. The horizontal axis denotes the coordinate of the one spatial direction ($x_\mu=0, \cdots, N_s-1$).The triangle (blue), circle (black) and plus (red) symbols denote the typical single configuration in vacuum in D, E and F in Fig.~\ref{fig:dis-Ploopt}, respectively. Three dotted lines (brown) denote $+2\pi/3$, $0$ and $-2\pi/3$ from top to bottom. }
\label{fig:dis-Ploopt-each-slice}
\end{center}
\end{figure}
Figure~\ref{fig:dis-Ploopt-each-slice} shows the value of $\theta(x_\mu)$ of the spatial slice average of the Polyakov loop for the one spatial direction ($x_\mu$)~\footnote{In Fig.~\ref{fig:dis-Ploopt-each-slice} we shift the label of the site for each configuration using the periodicity to show a domain-wall at the same position.}.
Here we show that the results for typical three types of configuration in the vacua D, E and F after the thermalization process.
Each color denotes the different configuration and corresponds to the one in Fig.~\ref{fig:dis-Ploopt}.
The horizontal axis denotes the coordinate of the one spatial direction.
That indicates that the configurations in the vacua D, E and F are the kink configurations connecting to the two different $Z_3$ vacua.
There are two ``center domains"~\cite{Asakawa:2012yv} where the $Z_3$ charge of the Polyakov loop in fifth direction is different each other.
Thus, the whole average of the Polyakov loop whose complex phase is $\pm \pi/3$ or $\pi$ complex shown in Fig.~\ref{fig:dis-Ploopt} is realized by the average of two different $Z_3$ clusters.

In Fig.~\ref{fig:dis-Ploopt-each-slice} although we show only the configurations where the volume of two domain is the same, the configurations where one domain is larger than the one of the complement are also generated in our simulation.
However, the latter configurations is unstable rather than the former one when we change the value of $\beta$.

The configuration with two domains whose volumes are the same does not change to the one with single domain even if we generate more than $100,000$ Sweeps although the typical autocorrelation length is a few hundred Sweeps.
We investigate the parameter region in which such configurations stably remain.
We carry out the simulation changing the value of $\beta$ using the configuration with two center domains as the initial configuration.
According to the phase structure in Fig.~\ref{fig:phase-L-8-T-4-gamma}, the critical $\beta$ of the fourth dimensional confined/deconfined phase transition is $\beta_c=5.70 \pm 0.10$, and the fifth dimensional phase transition occurs at $\beta_c=3.05 \pm 0.05$.
The configuration with two center domains remains only their intermediate region, namely PHASE (III).

\begin{figure}[h]
\begin{center}
\includegraphics[scale=0.45]{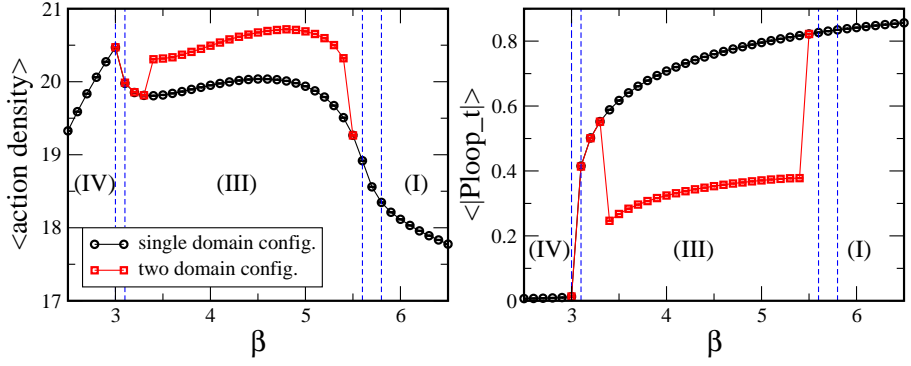}
\caption{Comparison of the action density (the magnitude of the Polyakov loop in the fifth direction) between the single center domain configurations and the two-domain configurations on the left (right) panel for $N_s=8, N_t=4, \gamma=3.00$. The double dotted (blue) lines in the lower and higher $\beta$ regions show the critical values of $\beta$ for the Ploop$_t$ and Ploop$_s$ shown in Fig.~\ref{fig:phase-L-8-T-4-gamma}. The notation (I), (III) and (IV) corresponds to the PHASE (I) -- (IV) in Fig.~\ref{fig:conj-phase-diagram}. }
\label{fig:Plaq-false-L8-T4}
\end{center}
\end{figure}

Figure~\ref{fig:Plaq-false-L8-T4} shows the comparisons between the single center domain configurations (circle (black) symbol) and the two center domains configurations (square (red) symbol) for the action density (left panel) and the magnitude of the Polyakov loop in the fifth direction (right panel).
The vacuum expectation value of the action density for the two-domain configurations is larger than the one for the single-domain configurations at the fixed $\beta$.

The difference of the action density can be explained as follows.
It comes from the interface of the domain.
The plaquette inside each domain is the same each other because of the $Z_3$ center symmetry of the action.
However the plaquette value in $\mu-5$ plane at the interface is different from the one inside each domain.
The difference between the plaquette value inside the domain and the one at the interface can be estimated as follows,
\beq
\Delta U_{\mu 5}(x) &=&U_{\mu}(x) U_{5}(x+\hat{\mu}a_4) U^\dag_\mu (x+\hat{5}a_5)U^\dag_5(x)\nonumber\\
&& - U_{\mu}(x) U_{5}(x+\hat{\mu}a_4) U^\dag_\mu (x+\hat{5}a_5)\tilde{U}^\dag_5(x),
\eeq
where the Polyakov loops for $U_5$ and $\tilde{U}_5$ have the different $Z_3$ charge each other.
The number of the plaquette in the interface is $N_s^3 N_t$.
Thus the action density should be proportional to $1/N_s$.

\begin{figure}[h]
\vspace{0.3cm}
\begin{center}
\includegraphics[scale=0.4]{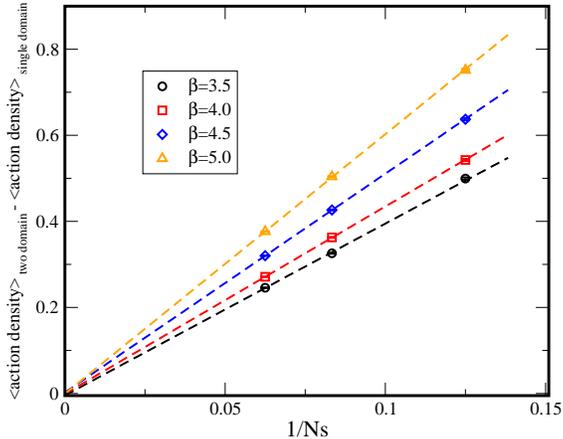}
\caption{Infinite volume limit for the difference of the action density between the single-domain configuration and the two-domain configuration at the fixed $N_t$, $\beta$ and $\gamma(=3.00)$. Each dotted line describes the linear extrapolation function in ($1/N_s$) for each fixed value of $\beta$.}
\label{fig:diff-action-false}
\end{center}
\end{figure}

Actually, we also investigate the four-dimensional volume dependence of the difference of the action density between the single-domain configurations and the two-domain configurations as shown in Fig.~\ref{fig:diff-action-false}.
The difference of the action density is proportional to $1/N_s$ as expected.

The appearance probability of the multi-domain configuration is the exponentially suppressed by the number of the interface, since the difference of the action density is also proportional to that.
In fact, we can not find the configurations with more than three domains when we generate configurations starting with more than $200$ independent random numbers.

The reasons why the configuration with the two-domain disappear in PHASE (I) would come from two origins, namely the exponential suppression of the appearance probability and
the phase transition in four-dimensional spaces.
In PHASE (I), the value of $\beta$ which is pre-factor of the plaquette in the action is larger than the one in PHASE (III).
The appearance rate of the configurations with mixed $Z_3$ is exponentially suppressed by the value of $\beta$.
Furthermore, in PHASE (I) the Polyakov loop in four-dimensional spaces also shows the deconfinement property, so that the link variables in the phase are ordered.
Then, the correlation for the plaquette in $\mu-5$ plane also becomes strong. 
It would make ordered the Polyakov loop in fifth direction for all spatial directions.
In the study on $3+1$ dimensional SU($3$) gauge theory~\cite{Borsanyi:2010cw,Endrodi:2014yaa}, the center clusters appear also in the deconfined phase with not extremely high temperature although the domain structure is not so clear in the $3+1$ dimensions.
It is a qualitatively similar region of $\beta$ if we consider that our lattice theory is a sort of the $4+1$ dimension finite temperature QCD.
We consider that the domain structure becomes clearly in $4+1$ dimensions since whose phase transition would be stronger than the one for $3+1$ dimensions because of a large number of bonds.

The similar situation occurs also in PHASE (II) with small $\gamma$, where the Polyakov loop in four-dimensional spaces shows the deconfinement property while
the one in fifth direction shows the confinement.
In fact, the two-domain configuration appears also in PHASE (II) in the small $\gamma$ region.
The strong anisotropy and a large volume to prevent the vacuum tunneling make such kink configuration with two domains.

\section{Summary and future directions}
We study the nonperturbative phase structure of the five-dimensional SU($3$) pure Yang-Mills theory on the lattice using the Wilson plaquette gauge action.
We perform the numerical simulation for the broad range of the lattice bare coupling constant and the anisotropy parameter.
To determine the phase structure, we observe the Polyakov loops in the fourth and the fifth directions.
Three types of the geometrical picture can be studied, namely the five-dimensional isotropic spaces, the $4+1$ dimensional geometry with the compactified fifth dimension and the one where the fifth lattice spacing is larger than the other dimensional lattice spacings.
Each geometry has a different phase structure.

In the isotropic five-dimensional case, there is a strong hysteresis between the deconfined phase and the confined phase in $4.2 \le \beta \le 4.5$ region.
The region does not depend on the lattice extent, so that we conclude it is the bulk first order phase transition.

If the fifth dimension is compact space, namely $(N_s a_4) > (N_t a_5)$, three phases appear.
A five-dimensional deconfined phase (PHASE (I)), a five-dimensional confined phase (PHASE (IV)) and
PHASE (III), where the fourth-dimensional Polyakov loop shows the confinement property, while the fifth-dimensional Polyakov loop shows the deconfinement.

On the other hand, if we take the naively opposite limit of the anisotropy parameter, namely we take $a_4 < a_5$, the different phase structure is obtained.
There are PHASE (I), (II) and (IV) in Fig.~\ref{fig:conj-phase-diagram}.
We determine that the critical values of the bare gauge coupling and the anisotropy parameter, $\gamma_c^{(s)}=0.70 \pm 0.10$ and $\beta=4.00 \pm 0.10$, in which the PHASE (II) shows up.
At first we considered that the appearance of the phase would be related with the ratio between the fourth dimensional physical length ($N_s a_4$) and the fifth-one ($N_t a_5$).
However, the critical values do not depend on the lattice extent in any directions and depend only on the value of $\gamma$.
It would suggest that the critical point depends only on the hierarchy of UV cutoff between fourth and fifth dimensions.
We expect that the layered brane geometry as expected by Fu and Nielsen~\cite{Fu:1984gj}, where each four-dimensional brane decouples with each other and all quark and gluon fields are localized on the brane nonperturbatively, is held in $\gamma \le \gamma_c^{(s)}$ region.
The localization of the gauge and matter fields is
necessary~\cite{Ohta:2010fu} to consider  ADD large extra dimension~\cite{ArkaniHamed:1998rs} and Randall-Sundrum~\cite{Randall:1999ee, Randall:1999vf} with the warped geometry models, since these models assume that only gravity can propagate the extra dimension.
The existence of PHASE(II) would support the theoretical possibility of such scenario.

Furthermore, we found that the appearance of the center domain-wall configuration in PHASE (II) and PHASE (III).
More than one center domain in single configuration where the $Z_3$ center charge are locally different coexists. 
Such configuration is a meta-stable state, since it does not change to the single domain configuration even if we continue the simulation more than $100,000$ Sweeps although its action density is larger than the one of the single domain configuration.

In this paper, we considered the phase structure only on the lattice.
However, it is interesting to consider to taking the continuum limit at least in four dimensions.
In both $\gamma \le \gamma_c^{(s)}$ and $\gamma_c^{(l)} \le \gamma$ regions, it is promising to define the continuum limit as the four-dimensional effective theory, since the strong hysteresis disappears and there is no bulk phase transition coming from the lattice artifact.

There are several recent studies on the five-dimensional SU($2$) gauge theory using the lattice simulations.
It is no difficulty to applying them for the SU($3$) gauge theory.
The following works have been done in the compactified fifth-dimensional lattice for the SU($2$) gauge theories.
In the low energy four-dimensional theory, we can obtain the gauge theory coupled to the adjoint scalar field (KK mode of $A_5$) whose mass is proportional to the inverse of the compactification radius.
The estimation of the scalar mass spectrum seems to be promising in the SU($2$) case~\cite{DelDebbio:2012mr}.
Introducing the orbifold boundary condition~\cite{Knechtli:2014ioa, Irges:2004gy, Irges:2012ih,Ishiyama:2009bk} is also interesting to construct the phenomenological model.
The boundary condition can explicitly break the gauge symmetry, and gives a rich phase structure in the low energy theory.
In particular, the SU($2$) gauge theory has a stick symmetry~\cite{Ishiyama:2009bk}, while the discussion cannot apply to the SU($3$) gauge theory.
The phase structure of the five-dimensional SU($3$) gauge theory with the orbifold boundary condition must be different from the SU($2$) case.
Moreover, we can introduce the finite temperature by introducing the another anisotropy between three-dimensional spaces and fourth dimension~\cite{Sakamoto:2007uy}.

The other important direction is to study the order of the phase transition around $\gamma = \gamma_c^{(s)}$.
In our simulation result, the value of $\gamma_c^{(s)}$ and the critical $\beta$ at the point are independent of the lattice extents both $N_s$ and $N_t$.
If the quark and gluon fields are localized on the brane nonperturbatively, then the four-dimensional branes are decoupled with each other~\cite{Fu:1984gj}.
Furthermore, it is suggested that the order of the phase transition at $\gamma_c^{(s)}$ changes to the second order one, since the point is the end point of the first phase transition in $\beta_4$--$\beta_5$ plane.
If there is the second order phase transition, then we can define a nontrivial continuum limit by tuning the value of the lattice parameter to the critical values.
The recent study on the five-dimensional SU($2$) lattice gauge theory shows the negative evidence of the second phase transition~\cite{DelDebbio:2013rka}.
We need a large lattice simulation to give a conclusion for the existence of the second order phase transition.

As a different direction, to study a multi-center-domain configuration would be interesting.
It is the first example to generate a clear (meta-)stable center domain configuration.
Practically, we can generate such configurations more than $100,000$ Sweeps, so that we can observe physical quantities using numerical simulation.
In particular, such multi-domain configuration would explain the phenomena of QGP~\cite{Asakawa:2012yv} in the finite temperature QCD. 
The measurements of the Polyakov loop correlator and the Wilson loop across the interface in $4+1$-dimensional SU($3$) lattice gauge theory would give a hint of qualitative properties of the center domains as a toy model of the finite temperature QCD.

Moreover, the improvement of the gauge action is the other direction~\cite{Kawai:1992um}.
The bulk phase transition, we observed in the isotropic case, comes from the lattice artifact.
The different lattice gauge action would avoid the artifact phase transition.
If it succeeds in avoiding the bulk phase transition even in the isotropic lattice, we would be able to study the criticality of the gauge theory in higher dimensions.

\section*{Acknowledgment}
We really appreciate P.~de~Forcrand and O.~Akerlund for carrying out the cross-check simulation and kindly telling us the
appearance of the center domains.
We also appreciate G.~Cossu carefully reading the draft and giving several important comments.
We also would like to thank M.~Kitazawa, T.~Iritani, M.~Sakamoto and S.~Ueda for useful comments and discussions.
The original numerical simulation code for four-dimensional PHB algorithm was provided by H.~Matsufuru.
We would like to appreciate his kindness.
We also would like to thank
A.~Irie for making several figures.
The work is inspired by the workshop on ``Toward Extra-dimension on the Lattice" at Osaka in March 2013.
Numerical simulation was carried out on Hitachi SR16000 at YITP, Kyoto University,
PC cluster at Kanazawa University,
NEC SX-8R and several PC clusters at RCNP, Osaka University.
We acknowledge Japan Lattice Data Grid for data transfer and storage.
E.I. is supported in part by Strategic Programs for Innovative Research (SPIRE) Field 5.
K.K. is supported by RIKEN Special Postdoctoral Researchers Program.


\newpage

\appendix

\section{Raw data}\label{sec:raw-data}

The raw data of the values of plaquette and Polyakov loop for $N_s=N_t=8$ with $\gamma=1.00$ are given in Table.~\ref{table:data-L-8-T-8-gamma-1.0}.

\begin{table}[h]
\begin{center}
\begin{tabular}{|c||c|c||c|c|}
\hline
\multicolumn{1}{|c||}{} &  \multicolumn{2}{c||}{Plaquette} & \multicolumn{2}{c|}{Polyakov loop} \\
\hline
$\beta$ &  cold start & hot start & cold start & hot start \\
\hline
$1.00$ & $0.060127(6 ) $  &  $ 0.060132 ( 10)$  & $4.60(3) \times 10^{-3}$  &   $ 4.63 (4) \times 10^{-3} $    \\          
$1.50$ & $0.093488(7 ) $  &  $ 0.093498 ( 9 )$  & $4.63(3) \times 10^{-3}$  &   $ 4.60 (3) \times 10^{-3} $    \\          
$2.00$ & $0.128917(8 ) $  &  $ 0.128900 ( 9 )$  & $4.65(3) \times 10^{-3}$  &   $ 4.61 (3) \times 10^{-3} $    \\             
$2.50$ & $0.166354(10) $  &  $ 0.166352 ( 8 )$  & $4.62(4) \times 10^{-3}$  &   $ 4.62 (3) \times 10^{-3} $    \\             
$3.00$ & $0.206120(9 ) $  &  $ 0.206109 ( 9 )$  & $4.61(4) \times 10^{-3}$  &   $ 4.63 (3) \times 10^{-3} $    \\             
$3.60$ & $0.258382(11) $  &  $ 0.258383 ( 12)$  & $4.69(3) \times 10^{-3}$  &   $ 4.63 (3) \times 10^{-3} $    \\           
$3.70$ & $0.267838(11) $  &  $ 0.267871 ( 12)$  & $4.66(4) \times 10^{-3}$  &   $ 4.61 (4) \times 10^{-3} $    \\           
$3.80$ & $0.277636(12) $  &  $ 0.277664 ( 12)$  & $4.66(3) \times 10^{-3}$  &   $ 4.65 (3) \times 10^{-3} $    \\           
$3.90$ & $0.287845(12) $  &  $ 0.287849 ( 12)$  & $4.64(3) \times 10^{-3}$  &   $ 4.66 (3) \times 10^{-3} $    \\           
$4.00$ & $0.298522(14) $  &  $ 0.298551 ( 14)$  & $4.66(3) \times 10^{-3}$  &   $ 4.63 (3) \times 10^{-3} $    \\           
$4.10$ & $0.309865(16) $  &  $ 0.309871 ( 13)$  & $4.64(4) \times 10^{-3}$  &   $ 4.68 (3) \times 10^{-3} $    \\           
$4.20$ & $0.485324(40) $  &  $ 0.322145 ( 14)$  & $3.49(11) \times 10^{-2}$ &   $ 4.62 (3) \times 10^{-3} $    \\           
$4.30$ & $0.515032(25) $  &  $ 0.335690 ( 17)$  & $5.29(16) \times 10^{-2}$ &   $ 4.60 (3) \times 10^{-3} $    \\           
$4.40$ & $0.536315(19) $  &  $ 0.351706 ( 23)$  & $6.93(17) \times 10^{-2}$ &   $ 4.59 (3) \times 10^{-3} $    \\           
$4.50$ & $0.554038(16) $  &  $ 0.374341 ( 41)$  & $8.12(22) \times 10^{-2}$ &   $ 4.60 (3) \times 10^{-3} $    \\           
$4.60$ & $0.569530(15) $  &  $ 0.569550 ( 16)$  & $9.17(28) \times 10^{-2}$ &   $ 9.45 (24) \times 10^{-2} $   \\           
$4.70$ & $0.583420(14) $  &  $ 0.583443 ( 13)$  & $1.04 (3) \times 10^{-1}$ &   $ 1.01 (3) \times 10^{-1} $    \\           
$4.80$ & $0.596102(15) $  &  $ 0.596092 ( 14)$  & $9.90(34) \times10^{-2}$  &   $ 9.74 (35) \times 10^{-2} $   \\           
$4.90$ & $0.607713(13) $  &  $ 0.607724 ( 13)$  & $1.29 (3) \times 10^{-1}$  &  $1.24 (3 ) \times 10^{-1}  $          \\    
$5.00$ & $0.618496(12) $  &  $ 0.618521 ( 13)$  & $1.29 (4) \times 10^{-1}$  &  $1.45 (3 ) \times 10^{-1}  $          \\    
$5.50$ & $0.663184(10) $  &  $ 0.663186 ( 10)$  & $1.94 (5) \times 10^{-1}$  &  $2.00 (3 ) \times 10^{-1}  $          \\      
$6.00$ & $0.697386(8 ) $  &  $ 0.697381 ( 8 )$  & $2.21 (7) \times 10^{-1}$  &  $2.46 (4 ) \times 10^{-1}  $          \\      
$6.50$ & $0.724768(7 ) $  &  $ 0.724773 ( 8 )$  & $2.79 (5) \times 10^{-1}$  &  $2.46 (8 ) \times 10^{-1}  $          \\      
$7.00$ & $0.747380(7 ) $  &  $ 0.747368 ( 7 )$  & $2.75 (9) \times 10^{-1}$  &  $3.11 (7 ) \times 10^{-1}  $          \\      
$7.50$ & $0.766398(6 ) $  &  $ 0.766389 ( 6 )$  & $3.21 (9) \times 10^{-1}$  &  $3.45 (7 ) \times 10^{-1}  $          \\      
$8.00$ & $0.782660(6 ) $  &  $ 0.782665 ( 5 )$  & $3.56 (7) \times 10^{-1}$  &  $3.96 (4 ) \times 10^{-1}  $          \\      
$8.50$ & $0.796755(4 ) $  &  $ 0.796754 ( 5 )$  & $4.23 (4) \times 10^{-1}$  &  $3.89 (9 ) \times 10^{-1}  $          \\      
$9.00$ & $0.809094(5 ) $  &  $ 0.809091 ( 4 )$  & $4.13 (9) \times 10^{-1}$  &  $3.88 (10) \times 10^{-1}  $ 	    \\    
$9.50$ & $0.819998(4 ) $  &  $ 0.820002 ( 5 )$  & $3.87 (1) \times 10^{-1}$  &  $4.67 (4 ) \times 10^{-1}  $       \\      
\hline
\end{tabular}
\caption{Plaquette and Polyakov loop for $N_s=N_t=8, \gamma=1.00$}
\label{table:data-L-8-T-8-gamma-1.0}
\end{center}
\end{table}

\newpage

\section{The phase diagrams for the larger lattice sizes}\label{sec:app-phase-diagram}
\begin{figure}[h]
\begin{center}
\includegraphics[scale=0.5]{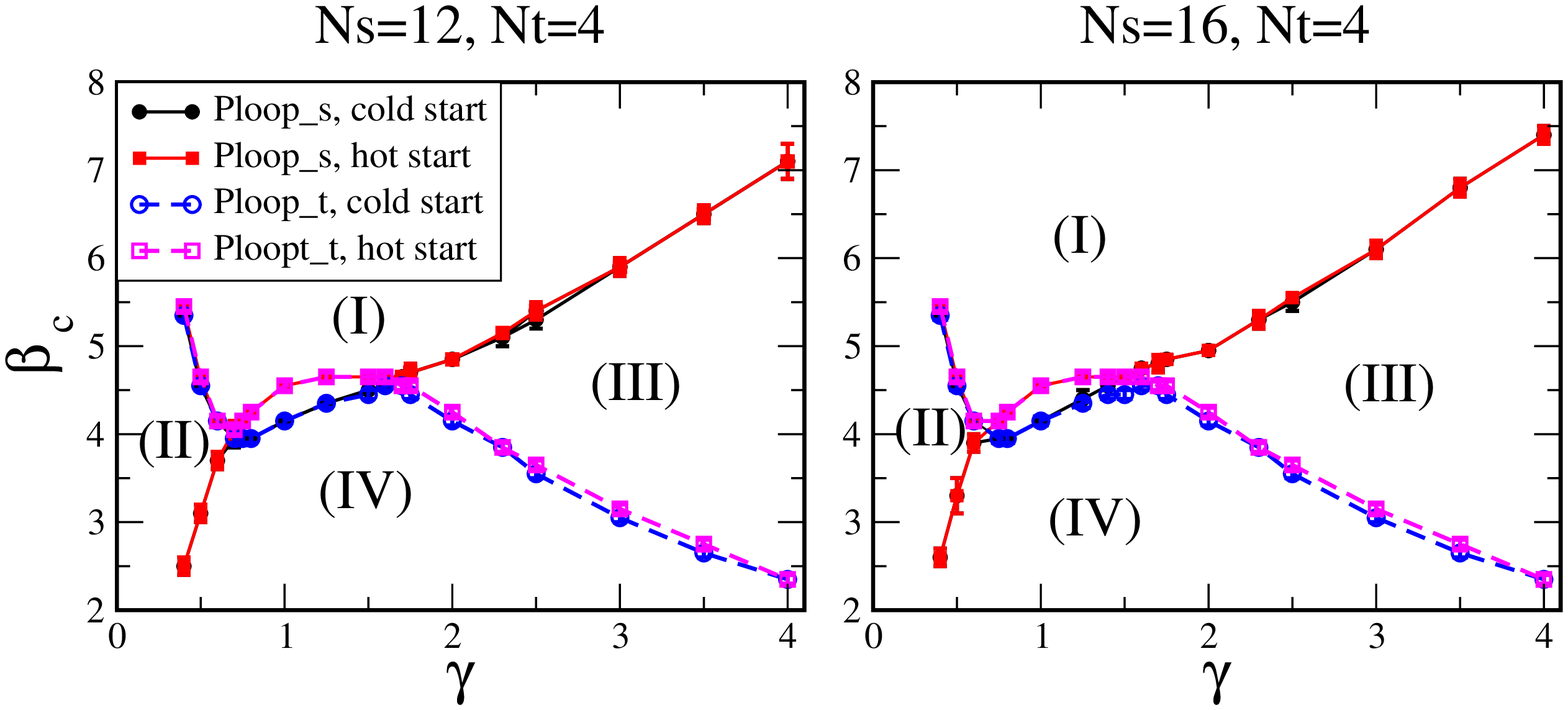}
\caption{Phase structure determined by $\beta_c$ of $\langle |$Ploop$_s| \rangle$ and $\langle |$Ploop$_t |\rangle$ for $N_s=12$ and $N_s=16$, $N_t=4$ lattices. }
\label{fig:phase-L-12-L-16-T-4-gamma}
\end{center}
\end{figure}

\begin{figure}[h]
\begin{center}
\includegraphics[scale=0.3]{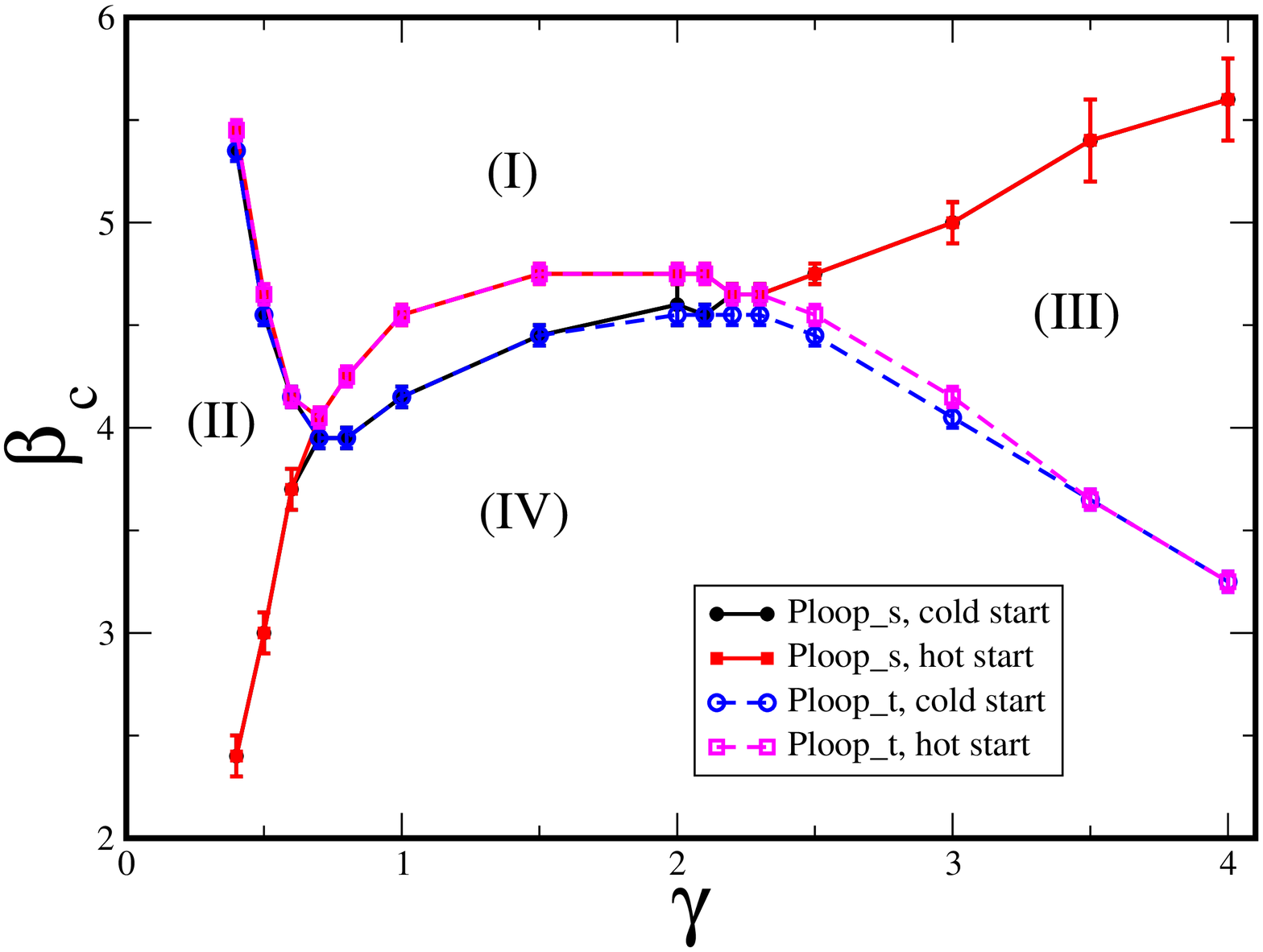}
\caption{Phase structure determined by $\beta_c$ of $\langle |$Ploop$_s | \rangle$ and $\langle |$Ploop$_t | \rangle$ for $N_s=12, N_t=6$ lattice. }
\label{fig:phase-L-12-T-6-gamma}
\end{center}
\end{figure}

We show that the phase diagrams on $\beta$--$\gamma$ plane for the larger lattice sizes.
Figure~\ref{fig:phase-L-12-L-16-T-4-gamma} and ~\ref{fig:phase-L-12-T-6-gamma} show the diagrams for $(N_s, N_t)=(12,4), (16,4)$ and $(12,6)$.

\end{document}